\documentclass[conference]{IEEEtran}
\IEEEoverridecommandlockouts

\usepackage{cite}
\usepackage{amsmath,amssymb,amsfonts}
\usepackage{algorithmic}
\usepackage{graphicx}
\usepackage{tikz}
\usepackage{textcomp}
\usepackage[table]{xcolor}
\def\BibTeX{{\rm B\kern-.05em{\sc i\kern-.025em b}\kern-.08em
    T\kern-.1667em\lower.7ex\hbox{E}\kern-.125emX}}

\usepackage{booktabs,multirow,tabularx,array,makecell,
    amsmath, amsthm}
\usepackage{listings}
\usepackage{subcaption}
\usepackage{todonotes}
\usepackage{tabularray}
\usepackage{etoolbox}
\usepackage{algorithm}
\usepackage{algorithmic}
\usepackage{caption}
\usepackage{hyperref}
\usepackage{enumitem}
\usepackage{xspace}

\newcommand{\fwName}{TurboFHE\xspace}

\begin{document}

\title{
TurboFHE: An Efficient and Scalable Multi-GPU Framework for Homomorphic LLM Inference

}

\author{
\IEEEauthorblockN{Honghui You\textsuperscript{1,2}\qquad
Haoping Yang\textsuperscript{1}\qquad
Wenzhe Wang\textsuperscript{2}\qquad
Lei Ju\textsuperscript{2,1\textdagger}}
\IEEEauthorblockA{\textsuperscript{1}\textit{School of Cyber Science and Technology, Shandong University}, Qingdao, China\\
\textsuperscript{2}\textit{Quan Cheng Laboratory}, Jinan, China\\
yhh21@mail.sdu.edu.cn\quad 202621183@mail.sdu.edu.cn\quad
ts-wangwzh@qcl.edu.cn\quad julei@sdu.edu.cn}
\thanks{\textsuperscript{\textdagger}Lei Ju is the corresponding author.}
}

\maketitle

\begin{abstract}
Fully homomorphic encryption (FHE) enables privacy-preserving large language model inference but remains computationally intensive. Existing multi-GPU FHE inference systems rely on expert-crafted parallelization plans that generalize poorly across hardware platforms and applications.
We present \fwName, a framework that automatically compiles CKKS inference programs into efficient multi-GPU executables. Its compiler can split different nodes of the same computation graph at the ciphertext and polynomial level. Guided by a measurement-based performance model, its hardware-aware planner jointly optimizes task partitioning, placement, operand residency, and transfer scheduling.
Its backend then generates optimized GPU kernels and executes plans using CUDA graphs and selective point-to-point transfers.
\fwName adapts its plans to diverse hardware configurations and FHE inference algorithms for large-scale models.
On a single A100 GPU, \fwName achieves a $24.5\times$ speedup over THOR's reported BERT-Base inference latency.
For Cerium's Llama-3-8B workload, \fwName outperforms Cerium's expert-crafted plans by $1.14\times$, $1.16\times$, and $1.24\times$ on 2, 4, and 8 RTX PRO~6000 GPUs, respectively.
On A100 GPUs, \fwName achieves $98.0\%$, $91.8\%$, and $79.3\%$ parallel efficiency for the same workload on 2, 4, and 8 GPUs, respectively.
\end{abstract}

\begin{IEEEkeywords}
    Privacy-Preserving LLM inference, Fully Homomorphic Encryption, Multi-GPU Acceleration
\end{IEEEkeywords}

\section{Introduction}
Fully homomorphic encryption (FHE) enables privacy-preserving
large language model (LLM) inference on sensitive personal or proprietary data~\cite{moon2025thor,park2026scaling}.
It allows cloud servers to compute directly on encrypted inputs and return encrypted outputs for clients to decrypt locally.
Among FHE schemes, CKKS is particularly well suited to LLM inference because it supports approximate arithmetic on SIMD-packed ciphertexts.

Unfortunately, FHE inference remains computationally intensive and memory demanding.
Prior work has accelerated core FHE primitives, including number-theoretic transforms (NTTs),
key switching, and bootstrapping, on
ASICs~\cite{kim2022bts,kim2022ark},
FPGAs~\cite{han2022coxhe,agrawal2023fab,yang2023poseidon},
and GPUs~\cite{jung2021over,fan2023tensorfhe,jiao2025neo,fan2025warpdrive}.
Among these platforms, commodity GPUs offer high throughput
and general-purpose programmability without custom
fabrication~\cite{kim2024cheddar},
enabling end-to-end encrypted inference for models such as CryptoNets~\cite{FGP}
BERT~\cite{moon2025thor} and Llama~\cite{park2026scaling}.
To further improve performance, recent studies have explored multi-GPU execution~\cite{aegis,jayashankar2026cerium,park2026scaling}, focusing primarily on high-end systems with fast interconnects or large GPU memory capacities.

Efficient multi-GPU FHE inference requires parallelization
plans that adapt to diverse hardware and evolving FHE algorithms.
However, manually tuned plans often fail to generalize across
these settings. Cerium, a state-of-the-art multi-GPU FHE
framework~\cite{jayashankar2026cerium}, relies on  programmers to manually specify partitions and the degree of parallelism for each stage.
Compared with a 4 GPUs deployment, Cerium's plans yield higher latency on 8 GPUs RTX~5090 work station across all tested benchmarks, due to restricted inter-GPU link bandwidth  (63\,GB/s) . Furthermore, algorithmic differences complicate parallelization.
We port THOR~\cite{moon2025thor}, a state-of-the-art design
for encrypted BERT inference, to Cerium.
On 8 NVSwitch-connected A100 GPUs, splitting every
FHE function across all GPUs at the polynomial granularity
achieves only a $2.48\times$ speedup.
Although several deep learning systems automate parallelization or memory planning~\cite{zheng2022alpa,jia2019flexflow,sheng2023flexgen},
adapting these techniques to minimize single-request latency in CKKS inference requires {jointly} optimizing partition granularity, communication, duplicated computation, and cryptographic constants reused across layers.

In general, the optimal multi-GPU deployment of an FHE LLM inference task is tightly coupled with the task model, FHE parameters, and hardware specifications (compute power, memory capacity, interconnect bandwidth, etc.). Manual tuning greatly increases the deployment barrier for the system due to the huge design space.
To this end, we introduce \fwName, a framework that
automatically generates parallelization plans for large-scale FHE
inference on multi-GPU systems.
Programmers specify only the homomorphic computation in a domain-specific language.
Given the program, a machine profile, and measured task costs,
the compiler jointly determines how to split computation, how to assign the resulting tasks to GPUs, how to supply their operands within device memory budget, and how to schedule tasks and data transfers.
The runtime instantiates the compiler-generated plans
as CUDA graphs and replays them for execution.
Our main contributions are as follows:
\begin{itemize}[leftmargin=*]
\item \textbf{Multi-GPU Performance model.}
Our model estimates plan latency using measured computation
and transfer costs.
It derives latency lower bounds from resource loads and
critical paths to prune candidates.
\item \textbf{Automatic multi-GPU program generation.}
The proposed compiler explores parallelization plans
for a give FHE LLM inference task.
Guided by the performance model, its hardware-aware planner
jointly optimizes task partitioning and placement across GPUs,
while coordinating operand residency and transfer scheduling
under memory constraints.

\item \textbf{Scalable FHE inference.}
We demonstrate automatic adaptation across FHE workloads and diverse consumer and datacenter GPU systems. On 8 RTX~5090 GPUs, \fwName completes BERT-Base inference in 8.1\,s.
For Cerium's Llama-3-8B implementation, it achieves $5.56$--$6.34\times$ speedups on eight GPUs across two platforms, outperforming Cerium by up to $1.24\times$.
\end{itemize}

\section{Background}
\label{sec:background}
\subsection{CKKS scheme}
CKKS~\cite{cheon2017homomorphic} packs real or complex vectors
into ciphertexts and supports approximate element-wise
addition and multiplication, as well as cyclic rotations.
Ciphertext polynomials are represented in a residue number
system (RNS) modulo a product of machine-word
primes~\cite{cheon2018full}.
An FHE program's multiplicative depth is the maximum number
of sequential multiplications along any dependency path.
In CKKS, rescaling after multiplication reduces the ciphertext scale by removing one or more prime factors from its modulus.
With fixed scaling parameters and prime sizes, the number of further multiplications a ciphertext can support grows roughly linearly with the remaining prime count.
Bootstrapping restores the ciphertext modulus to enable further computation~\cite{cheon2018bootstrapping}.
Homomorphic operations are implemented using number-theoretic
transforms (NTTs), pointwise modular arithmetic, and RNS basis
conversion.
Rotations and ciphertext multiplication each involves a key switch,
which extends the ciphertext to a larger basis, takes its inner product with an evaluation key,
and converts the result back to the original basis.
NTTs and pointwise arithmetic operate independently per polynomial.
Basis conversion combines residues across polynomials to change
the set of primes, requiring communication when these polynomials
reside on different GPUs.
FHE inference also requires evaluation keys, encoded plaintext
constants, and temporary workspace, whose sizes and concurrent
residency affect GPU memory usage.

\subsection{Motivation}
\label{sec:motivation}
\begin{table}[t]
\centering
\caption{Multi-GPU speedup ($S_P=T_1/T_P$) with Cerium's
backend under different hardware, homomorphic programs,
and parallelization plans.}
\label{tab:motivation-cerium}
\setlength{\tabcolsep}{6pt}
\begin{tabular}{@{}lrrrr@{}}
\toprule
\multicolumn{5}{@{}l}{\textbf{(a) Fixed bootstrapping program and plan}} \\
GPU / link & $T_1$ & $S_2$ & $S_4$ & $S_8$ \\
\midrule
H100 / NVSwitch\textsuperscript{a}
    & 19.5\,ms & 1.42 & 1.84 & 2.01 \\
H100 / NVL pairs\textsuperscript{b}
    & 22.4\,ms & 1.56 & 1.31 & -- \\
RTX 5090 / PCIe
    & 24.5\,ms & 1.49 & 1.54 & 1.34 \\
\midrule
\multicolumn{5}{@{}l}{\textbf{(b) BERT-Base (A100/NVSwitch)}} \\
Program / plan & $T_1$ & $S_2$ & $S_4$ & $S_8$ \\
\midrule
Cerium / original\textsuperscript{a}
    & 5.51\,s & 1.49 & 2.69 & 3.37 \\
Cerium / all-GPU\textsuperscript{c}
    & 5.60\,s & 1.44 & 1.77 & 1.90 \\
THOR / adapted\textsuperscript{d}
    & 5.16\,s & 1.54 & 1.83 & 1.96 \\
THOR / all-GPU\textsuperscript{c}
    & 5.16\,s & 1.58 & 2.11 & 2.48 \\
\bottomrule
\end{tabular}

\vspace{1pt}
\begin{minipage}{\columnwidth}
\scriptsize
\textsuperscript{a}Cerium's reported results~\cite{jayashankar2026cerium}; BERT latencies are divided by 12 for per-layer averages.
All other rows are our measurements.
\textsuperscript{b}4 H100 NVL GPUs.
\textsuperscript{c}All FHE functions use all $P$ GPUs
with polynomial-level parallelization.
\textsuperscript{d}THOR~\cite{moon2025thor} is ported to
Cerium; the adapted plan follows Cerium's BERT partition.
\end{minipage}
\end{table}

To examine how hardware and homomorphic algorithms affect
multi-GPU scalability, we use Cerium~\cite{jayashankar2026cerium},
an open-source, state-of-the-art multi-GPU CKKS framework,
as a common execution framework (Table~\ref{tab:motivation-cerium}).
Cerium lets programmers specify GPU counts, computation
partitions, and data transfers for FHE functions, while its
compiler generates polynomial-level GPU kernels and
communication schedules.

In (a), we run the same bootstrapping program and parallel
plan across different GPU systems.
Scaling from 2 to 4 H100s increases speedup from
$1.42\times$ to $1.84\times$ on NVSwitch, but reduces it
from $1.56\times$ to $1.31\times$ when communication
crosses NVLink pairs over PCIe.
Parallelization still provides modest gains over PCIe:
4 RTX~5090s achieve a $1.54\times$ speedup,
which drops to $1.34\times$ on 8 GPUs.

Thus, using more GPUs does not necessarily reduce latency.

In (b), we compare two homomorphic BERT-Base implementations:
Cerium's released implementation and
THOR~\cite{moon2025thor}, which uses different algorithms
for encrypted matrix operations and nonlinearities.
We port THOR to Cerium and evaluate two parallelization plans:
(i)~\emph{all-GPU}, which partitions every FHE function
across all $P$ GPUs at the RNS-polynomial level; and
(ii)~\emph{original/adapted}, which retains Cerium's original
BERT plan for Cerium's BERT implementation and adapts its plan to THOR.
Both use Cerium's compiler and runtime.
On A100/NVSwitch, the all-GPU plan achieves a 1.90$\times$ speedup,
compared with the 3.37$\times$ reported for Cerium's
original plan.
but improves THOR's from $1.96\times$ to $2.48\times$.
Thus, the preferred parallel plan depends on the
homomorphic implementation, even for the same model and hardware.

\section{Design}
\label{sec:overview}
\fwName records a CKKS inference program as a graph of ciphertext
operations and their dependencies (Figure~\ref{fig:overview}).
The planner uses measured computation and transfer costs to decide
how to split graph nodes into tasks and where to run each task.
It also tests whether rewriting regions on the critical path would
reduce latency. The compiler translates the selected plan into a
program for each GPU specifying its computations and data transfers.

\begin{figure}[t]
\centering
\input{figures/workflow}
\caption{\fwName workflow.}
\label{fig:overview}
\end{figure}

\subsection{Programming Interface}
\label{sec:dsl}

The \fwName DSL expresses neural-network operations, such as softmax
and feed-forward blocks, as compositions of CKKS operations in Python.
Programs written in the DSL specify neither GPU parallelism nor data transfers.

\noindent\textbf{Running example.}
We plan THOR's initial attention computation for 2 GPUs
(Figure~\ref{fig:example}). It forms 64 rotated ciphertexts from 4
input for the query, key, and value projections, then computes
and bootstraps 4 attention-score ciphertexts.

\begin{figure}[t]
\input{figures/example-front}
\caption{THOR's initial attention operations in the DSL, condensed
from its public implementation~\cite{moon2025thor}.
\texttt{dot\_pt} sums plaintext products over the rotated ciphertexts,
\texttt{align} aligns slots, and \texttt{scores} denotes the
query--key product.}
\label{fig:example}
\end{figure}

\subsection{Computation Graph and Partitioning}
\label{sec:graph}
A profiling run with the recording evaluator converts a DSL
program into a computation graph of ciphertext operations and
their dependencies.
Initially, each bootstrap forms a node, while other ciphertext
operations are grouped into \textbf{regions}.
Applications may define regions around neural-network operations;
the rest form maximal connected regions bounded
by bootstraps.
Each edge records a producer--consumer dependency and the
byte size of the corresponding value.
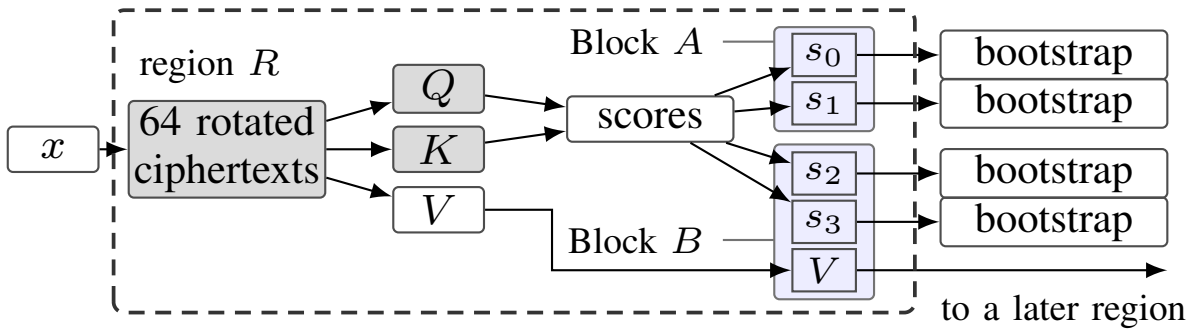
\begin{figure}[t]
\centering
\begin{tikzpicture}[
  x=1cm,y=0.8cm,
  op/.style={draw=black!70,rounded corners=1pt,minimum width=6mm,
             minimum height=3mm,inner sep=1pt,font=\footnotesize},
  pre/.style={op,fill=black!14},
  blk/.style={draw=black!55,fill=blue!7,rounded corners=1pt},
  res/.style={draw=black!70,minimum width=4.2mm,minimum height=2.8mm,
              inner sep=0.8pt,font=\scriptsize},
  title/.style={anchor=west,font=\footnotesize\bfseries},
  note/.style={font=\scriptsize},
  dep/.style={->,>=latex,thin},
  bar/.style={draw=black!70,minimum height=3.6mm,inner sep=0pt,
              font=\scriptsize,anchor=west},
  comp/.style={bar,fill=blue!10},
  xchg/.style={bar,fill=black!25},
  wait/.style={bar,fill=white,densely dotted}
]

\draw[blk] (5.17,4.90) rectangle (5.83,5.76);
\draw[blk] (5.17,3.51) rectangle (5.83,4.79);
\draw[thin,draw=black!55] (5.17,5.64) -- (4.83,5.64) node[note,left] {Block $A$};
\draw[thin,draw=black!55] (5.17,4.00) -- (4.83,4.00) node[note,left] {Block $B$};
\node[op] (x) at (0.40,4.75) {$x$};
\node[pre,minimum width=13mm,align=center] (cp) at (1.55,4.75) {64 rotated\\ciphertexts};
\node[pre] (q) at (2.95,5.25) {$Q$};
\node[pre] (k) at (2.95,4.75) {$K$};
\node[op] (v) at (2.95,4.25) {$V$};
\node[op,minimum width=11mm] (sc) at (4.35,5.00) {scores};
\node[res] (s0) at (5.50,5.53) {$s_0$};
\node[res] (s1) at (5.50,5.13) {$s_1$};
\node[res] (s2) at (5.50,4.55) {$s_2$};
\node[res] (s3) at (5.50,4.15) {$s_3$};
\node[res] (vo) at (5.50,3.75) {$V$};
\node[op,minimum width=15mm] (r0) at (7.02,5.53) {bootstrap};
\node[op,minimum width=15mm] (r1) at (7.02,5.13) {bootstrap};
\node[op,minimum width=15mm] (r2) at (7.02,4.55) {bootstrap};
\node[op,minimum width=15mm] (r3) at (7.02,4.15) {bootstrap};
\draw[dep] (x) -- (cp);
\draw[dep] (cp) -- (q);
\draw[dep] (cp) -- (k);
\draw[dep] (cp) -- (v);
\draw[dep] (q) -- (sc);
\draw[dep] (k) -- (sc);
\draw[dep] (sc) -- (s0);
\draw[dep] (sc) -- (s1);
\draw[dep] (sc) -- (s2);
\draw[dep] (sc) -- (s3);
\draw[dep] (s0) -- (r0);
\draw[dep] (s1) -- (r1);
\draw[dep] (s2) -- (r2);
\draw[dep] (s3) -- (r3);
\draw[dep] (v) -- (3.70,4.25) -- (3.70,3.75) -- (vo);
\draw[dep] (vo) -- (7.77,3.75);
\node[note,anchor=north west] at (6.15,3.70) {to a later region};
\draw[densely dashed,draw=black!80,line width=0.7pt,rounded corners=2pt] (0.80,3.40) rectangle (6.10,5.90);
\node[note,anchor=west] at (0.85,5.44) {region $R$};

\end{tikzpicture}
\caption{Computation graph of the running example.
Region $R$ (dashed box) produces score blocks $A$ and $B$,
followed by 4 bootstraps.
$Q$, $K$, and $V$ denote projection sums over rotated
ciphertexts; shaded operations are shared by both blocks.}
\label{fig:example-graph}
\end{figure}

\noindent\textbf{Granularity and splitting.}A \textbf{task} is a unit of work assigned to a single GPU.
Operations within each task are lowered to GPU kernel
invocations, which are indivisible execution units.
The graph supports splitting at two levels:
ciphertext operations and residue-polynomial operations.
Graph nodes can remain intact or be split into tasks at two levels.
\emph{Ciphertext-level splitting} distributes a node's output
ciphertexts across tasks, potentially duplicating shared
intermediate operations.
\emph{Polynomial-level splitting}, as in
Cinnamon~\cite{jayashankar2025cinnamon} and
Cerium~\cite{jayashankar2026cerium}, distributes the residue
polynomials of each ciphertext across tasks.
This exposes parallelism within a ciphertext but requires
cross-GPU communication at basis conversions.

\begin{figure}[t]
\centering
\begin{tikzpicture}[
  x=1cm,y=1cm,
  task/.style={draw=black!70,rounded corners=1.5pt,inner sep=0pt},
  region/.style={draw=black!80,line width=0.7pt,densely dashed,rounded corners=2pt,inner sep=0pt},
  cell/.style={draw=black!55,fill=blue!7,inner sep=0pt,font=\scriptsize},
  polynomial/.style={cell,fill=orange!25},
  shared/.style={draw=black!55,fill=black!14,inner sep=0pt,font=\scriptsize},
  res/.style={draw=black!70,inner sep=0pt,font=\scriptsize},
  title/.style={anchor=west,font=\footnotesize\bfseries},
  note/.style={font=\scriptsize},
  dep/.style={->,>=latex,thin}
]
\path[use as bounding box] (-0.05,0.12) rectangle (8.35,4.22);

\node[title] at (0.000,4.020) {(a) Ciphertext-level splitting};
\node[note] at (1.900,2.200) {shaded operations run once per task};
\node[res,minimum width=0.40cm,minimum height=0.36cm] at (0.200,3.100) {$x$};
\node[task,minimum width=1.70cm,minimum height=0.56cm] at (1.600,3.520) {};
\node[shared,align=center,minimum width=0.95cm] at (1.330,3.520) {rotations\\[-1pt]$Q$, $K$};
\node[cell,minimum width=0.55cm,minimum height=0.36cm] at (2.130,3.520) {$A$};
\draw[dep] (0.400,3.180) -- (0.750,3.520);
\draw[dep] (2.450,3.520) -- (2.730,3.520);
\node[note,anchor=west] at (2.780,3.520) {$s_0$, $s_1$};
\node[task,minimum width=1.70cm,minimum height=0.56cm] at (1.600,2.680) {};
\node[shared,align=center,minimum width=0.95cm] at (1.330,2.680) {rotations\\[-1pt]$Q$, $K$};
\node[cell,minimum width=0.55cm,minimum height=0.36cm] at (2.130,2.680) {$B$};
\draw[dep] (0.400,3.020) -- (0.750,2.680);
\draw[dep] (2.450,2.680) -- (2.730,2.680);
\node[note,anchor=west] at (2.780,2.680) {$s_2$, $s_3$, $V$};

\node[title] at (4.250,4.020) {(b) Polynomial-level splitting};
\node[note] at (6.150,2.200) {exchange at each basis conversion};
\node[res,minimum width=0.40cm,minimum height=0.36cm] at (4.450,3.100) {$s_0$};
\node[task,minimum width=2.40cm,minimum height=0.56cm] at (6.080,3.520) {};
\node[polynomial,minimum width=2.20cm,minimum height=0.36cm] at (6.080,3.520) {polynomials 1, 2};
\draw[dep] (4.650,3.180) -- (4.880,3.480);
\draw[dep] (7.280,3.480) -- (7.480,3.180);
\node[task,minimum width=2.40cm,minimum height=0.56cm] at (6.080,2.680) {};
\node[polynomial,minimum width=2.20cm,minimum height=0.36cm] at (6.080,2.680) {polynomials 3, 4};
\draw[dep] (4.650,3.020) -- (4.880,2.720);
\draw[dep] (7.280,2.720) -- (7.480,3.020);
\draw[<->,>=latex,thin] (6.080,3.240) -- (6.080,2.960);
\node[note,anchor=west] at (7.480,3.100) {result};

\node[title] at (0.000,1.770) {(c) Extract shared operations};
\node[note] at (2.000,0.700) {both blocks reuse the results};
\node[res,minimum width=0.40cm,minimum height=0.36cm] at (0.200,1.220) {$x$};
\node[region,minimum width=1.15cm,minimum height=0.60cm] at (1.250,1.220) {};
\node[shared,align=center,minimum width=0.95cm] at (1.250,1.220) {rotations\\[-1pt]$Q$, $K$};
\node[region,minimum width=1.50cm,minimum height=0.60cm] at (2.950,1.220) {};
\node[cell,minimum width=0.55cm,minimum height=0.36cm] at (2.620,1.220) {$A$};
\node[cell,minimum width=0.55cm,minimum height=0.36cm] at (3.280,1.220) {$B$};
\draw[dep] (0.400,1.220) -- (0.675,1.220);
\draw[dep] (1.825,1.220) -- (2.200,1.220);
\draw[dep] (3.700,1.220) -- (4.000,1.220);

\node[title] at (4.250,1.770) {(d) Rewrite: sums by terms};
\node[note] at (6.250,0.700) {partial sums, then added};
\node[res,minimum width=0.40cm,minimum height=0.36cm] at (4.450,1.220) {$x$};
\node[region,minimum width=1.40cm,minimum height=0.60cm] at (5.500,1.220) {};
\node[shared,minimum width=0.52cm,minimum height=0.36cm] at (5.180,1.220) {$\Sigma_1$};
\node[shared,minimum width=0.52cm,minimum height=0.36cm] at (5.820,1.220) {$\Sigma_2$};
\node[region,minimum width=1.70cm,minimum height=0.60cm] at (7.200,1.220) {};
\node[shared,minimum width=0.34cm,minimum height=0.36cm] at (6.630,1.220) {$+$};
\node[cell,minimum width=0.50cm,minimum height=0.36cm] at (7.130,1.220) {$A$};
\node[cell,minimum width=0.50cm,minimum height=0.36cm] at (7.710,1.220) {$B$};
\draw[dep] (4.650,1.220) -- (4.800,1.220);
\draw[dep] (6.200,1.220) -- (6.350,1.220);
\draw[dep] (8.050,1.220) -- (8.300,1.220);

\node[task,minimum width=0.30cm,minimum height=0.24cm] at (0.150,0.320) {};
\node[note,anchor=west] at (0.360,0.320) {task (one GPU)};
\node[region,minimum width=0.30cm,minimum height=0.24cm] at (2.750,0.320) {};
\node[note,anchor=west] at (2.960,0.320) {region};
\node[shared,minimum width=0.30cm,minimum height=0.24cm] at (4.350,0.320) {};
\node[note,anchor=west] at (4.560,0.320) {work both blocks need};
\end{tikzpicture}
\caption{Splitting and rewriting the running example:
(a) ciphertext-level splitting of $R$;
(b) polynomial-level splitting of a bootstrap;
(c) shared-computation extraction;
(d) homomorphic-sum restructuring.
GPU placement is determined separately.}
\label{fig:example-partition}
\end{figure}
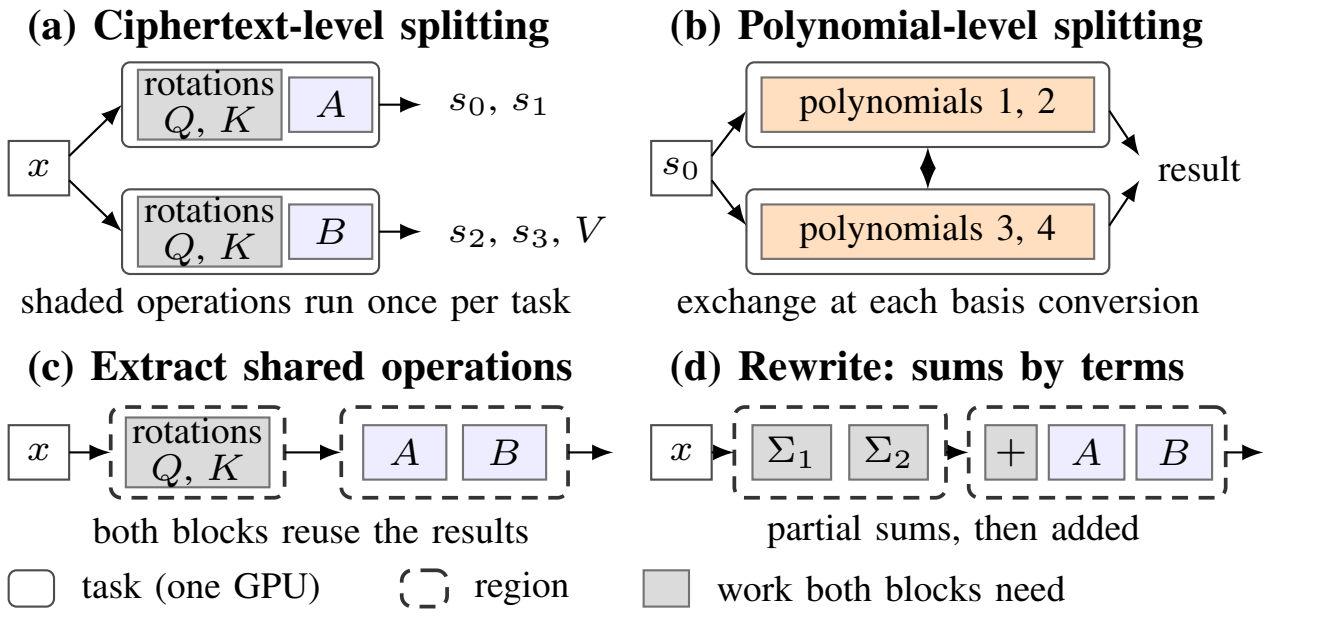

\noindent\textbf{Rewriting.}
Unlike splitting, a \textbf{rewrite} replaces a region with
multiple regions that collectively preserve its outputs.
Each resulting region can remain intact, be split, or undergo
further rewrites.
We use two rewrites.
\emph{Shared-computation extraction} moves operations that
would otherwise be duplicated across tasks into a shared region
computed once.
\emph{Sum restructuring} divides a homomorphic sum into
grouped partial sums and a separate region that combines them.
The partial-sum region can then be split by group.

\noindent\textbf{Multi-level optimizations.}
We also optimize computation at multiple levels.
At the ciphertext level, we defer rescaling where valid.
At the polynomial level, common-subexpression elimination
shares key-switch intermediates across rotations of the same
ciphertext (hoisting~\cite{bossuat2021efficient,halevi2018faster})
and moves final basis conversions past additions when valid.
Furthermore, rotations with compatible key-switch
structures and plaintext encodings are batched into fewer,
larger kernel invocations.

\noindent\textbf{Partition.}
A \textbf{partition} specifies graph rewrites and the
decomposition of each resulting node into one or more tasks,
including the splitting granularity and distribution of work
among tasks.
These decisions determine the tasks and their dependencies.
Each task records its required evaluation keys and plaintext
constants, collectively called \emph{operands}.
Cross-GPU transfers occur only between tasks.

\noindent\textbf{Example.}
Figure~\ref{fig:example-partition} illustrates these splitting and rewriting choices.
In (a), splitting $R$ by output block duplicates input rotations
and the $Q$ and $K$ projections across two tasks.
In (b), polynomial-level splitting assigns half the residue
polynomials to each task, requiring exchanges at basis conversions.
In (c), extracting shared rotations and projections avoids
duplication, but placing $V$ on another GPU requires transferring
all 64 rotated ciphertexts.
In (d), grouping projection terms by their rotated ciphertexts
lets tasks generate only the required rotations and partial sums,
avoiding redundant computation and transfers.
A second region combines these sums before computing the scores.

\subsection{Plans}
\label{sec:plans}
A \textbf{plan} determines how this graph is
executed across GPUs through four decisions.
\textbf{Partition} divides graph computations into tasks;
\textbf{placement} assigns each task to a GPU;
\textbf{schedule} orders tasks and transfers;
and \textbf{residency} determines which operands remain in GPU
memory or are streamed from the host and whether plaintext constants
are compact or encoded.
The objective is to minimize single-request latency $T$.

\subsection{Performance Model}
\label{sec:model}

\noindent\textbf{Cost calibration.}
Before plan search, we benchmark graph nodes and their operations.
Intact nodes use their measured times.
For ciphertext-level splitting and rewrites, task times are
estimated from the original node's measured time, scaled by
the assigned operations' relative costs.
These ratios may sum to more than one due to duplicated work.
Tasks at the polynomial level are benchmarked directly.
A machine profile records pairwise GPU transfer costs
by data size and per-GPU host-upload rates.
Communication patterns are modeled as compositions
of point-to-point transfers.

\noindent\textbf{Simulation.}
Using these costs, the model simulates task execution and transfers.
Tasks execute when their inputs and operands
are ready, while uploads for subsequent computations
may overlap ongoing execution.
Predicted latency $\hat{T}$ is the  completion time of the final task.
The critical path identifies the tasks, transfers, and waits
determining this latency.

\noindent\textbf{Lower bound.}
A {partial plan} only fixes partition and placement; a {full plan}
also fixes schedule and residency.
For pruning, we derive a lower bound on predicted latency
under the calibrated cost model for any full plan extending
a partial plan.
It is the maximum of the busiest resource's load (GPU, link port,
or host path) and the longest dependency path including communication
delays~\cite{papadimitriou1988towards}.
The dependency path includes mandatory per-GPU ordering
constraints and minimum communication delays.
The bound assumes optimal scheduling and residency,
counting only unavoidable uploads.

\subsection{Planner}
\label{sec:planner}
\noindent\textbf{Partitioning and placement search.}
Starting from multiple placements, coordinate descent~\cite{wright2015coordinate}
tests splits and task placements for one node at a time, keeping the
others fixed. The planner prunes a candidate if its lower bound $L$
cannot improve on the best feasible plan's predicted latency. For the
remaining candidates, it derives a schedule and keeps the choice with
the lowest predicted latency $\hat{T}$.
When a full pass makes no changes, the planner tests the 2 rewrites
in \S\ref{sec:graph} on regions along the best plan's critical path,
starting from the current placement. An improving rewrite restarts
coordinate descent. Otherwise, the search proceeds to final selection.

\noindent\textbf{Scheduling.}
The planner batches transfers into logical rounds without global
barriers. By default, a GPU sends at round end. Early sends can overlap
the remaining computation once all values in a batch are ready and
transfer order permits. A consistent transfer order prevents deadlock.
Only the GPUs participating in a transfer synchronize.

\noindent\textbf{Residency and final selection.}
For the best-ranked plans, exact branch and bound selects
evaluation-key sets to stream, minimizing added latency
under GPU memory constraints.
Plaintext constants are stored as rearranged vectors or
single-period representations~\cite{You2026MACH,shi2025wpc}.
Spare GPU memory caches encoded forms according to
latency saved per byte.
Streamed operands are released between distant uses and
prefetched before reuse; tasks wait for missing operands.
Finally, we compile and benchmark tasks for the
top-ranked candidate plans, refine their latency estimates,
and select the plan with the lowest updated $\hat{T}$.

\begin{figure}[t]
\centering
\begin{tikzpicture}[
  x=1cm,y=1cm,
  bar/.style={draw=black!70,inner sep=0pt,font=\scriptsize},
  comp/.style={bar,fill=blue!10},
  xchg/.style={bar,fill=black!30},
  wait/.style={bar,fill=white,densely dotted,text=black!60},
  upl/.style={bar,fill=black!12},
  note/.style={font=\scriptsize},
  title/.style={anchor=west,font=\footnotesize\bfseries},
  region/.style={draw=black!80,line width=0.7pt,densely dashed,rounded corners=2pt,inner sep=0pt},
  res/.style={draw=black!70,inner sep=0pt,font=\scriptsize},
  dep/.style={->,>=latex,thin}
]
\path[use as bounding box] (-0.05,0.18) rectangle (8.35,5.28);

\node[title] at (-0.05,5.10) {(a) Partition and placement};
\node[note,anchor=east] at (1.150,4.380) {GPU 0};
\node[note,anchor=east] at (1.150,3.800) {GPU 1};
\node[note] at (2.450,4.810) {partial sums};
\node[note] at (4.150,4.810) {add};
\node[note] at (6.000,4.810) {bootstraps};
\node[res,minimum width=0.36cm,minimum height=0.36cm] at (1.450,4.380) {$x$};
\node[region,minimum width=1.20cm,minimum height=1.12cm] at (2.450,4.090) {};
\node[comp,minimum width=0.90cm,minimum height=0.36cm] at (2.450,4.380) {$\Sigma_1$};
\node[comp,minimum width=0.90cm,minimum height=0.36cm] at (2.450,3.800) {$\Sigma_2$};
\node[region,minimum width=1.60cm,minimum height=0.54cm] at (4.150,4.380) {};
\node[comp,minimum width=1.30cm,minimum height=0.36cm] at (4.150,4.380) {$+$, scores};
\node[comp,minimum width=0.50cm,minimum height=0.36cm] at (5.700,4.380) {$s_0$};
\node[comp,minimum width=0.50cm,minimum height=0.36cm] at (6.300,4.380) {$s_1$};
\node[comp,minimum width=0.50cm,minimum height=0.36cm] at (5.700,3.800) {$s_2$};
\node[comp,minimum width=0.50cm,minimum height=0.36cm] at (6.300,3.800) {$s_3$};
\draw[dep] (1.630,4.380) -- (2.000,4.380);
\draw[dep] (1.600,4.200) -- (2.000,3.840);
\draw[dep] (2.900,4.380) -- (3.500,4.380);
\draw[dep] (2.900,3.800) -- (3.500,4.260);
\draw[dep] (4.800,4.380) -- (5.450,4.380);
\draw[dep] (4.800,4.260) -- (5.450,3.840);
\draw[dep] (6.550,4.380) -- (7.250,4.380);
\draw[dep] (6.550,3.800) -- (7.250,4.300);
\node[note,anchor=west] at (7.280,4.380) {results};

\node[title] at (-0.05,3.300) {(b) Execution};

\node[note,anchor=east] at (1.150,2.450) {GPU 0};
\node[note,anchor=east] at (1.150,2.030) {host link};
\node[note,anchor=east] at (1.150,1.250) {GPU 1};
\node[note,anchor=east] at (1.150,0.830) {host link};
\draw[black!25] (1.25,2.450) -- (8.30,2.450);
\draw[black!25] (1.25,2.030) -- (8.30,2.030);
\draw[black!25] (1.25,1.250) -- (8.30,1.250);
\draw[black!25] (1.25,0.830) -- (8.30,0.830);

\node[xchg,minimum width=0.25cm,minimum height=0.40cm] at (1.375,2.450) {};
\node[xchg,minimum width=0.25cm,minimum height=0.40cm] at (1.375,1.250) {};
\draw[dep] (1.375,2.250) -- (1.375,1.450);
\node[note] at (1.375,2.870) {$x$};
\node[xchg,minimum width=0.28cm,minimum height=0.40cm] at (2.740,2.450) {};
\node[xchg,minimum width=0.28cm,minimum height=0.40cm] at (2.740,1.250) {};
\draw[dep] (2.740,1.450) -- (2.740,2.250);
\node[note] at (2.740,2.870) {sums};
\node[xchg,minimum width=0.26cm,minimum height=0.40cm] at (4.310,2.450) {};
\node[xchg,minimum width=0.26cm,minimum height=0.40cm] at (4.310,1.250) {};
\draw[dep] (4.310,2.250) -- (4.310,1.450);
\node[note] at (4.310,2.870) {$s_2$, $s_3$};
\node[xchg,minimum width=0.26cm,minimum height=0.40cm] at (8.070,2.450) {};
\node[xchg,minimum width=0.26cm,minimum height=0.40cm] at (8.070,1.250) {};
\draw[dep] (8.070,1.450) -- (8.070,2.250);
\node[note,anchor=east] at (8.300,2.870) {results};

\node[comp,minimum width=1.10cm,minimum height=0.40cm] at (2.050,2.450) {$\Sigma_1$};
\node[comp,minimum width=1.30cm,minimum height=0.40cm] at (3.530,2.450) {$+$, scores};
\node[comp,minimum width=1.75cm,minimum height=0.40cm] at (5.315,2.450) {bootstrap $s_0$};
\node[comp,minimum width=1.75cm,minimum height=0.40cm] at (7.065,2.450) {bootstrap $s_1$};

\node[comp,minimum width=1.10cm,minimum height=0.40cm] at (2.050,1.250) {$\Sigma_2$};
\node[wait,minimum width=1.30cm,minimum height=0.40cm] at (3.530,1.250) {wait};
\node[comp,minimum width=1.75cm,minimum height=0.40cm] at (5.315,1.250) {bootstrap $s_2$};
\node[comp,minimum width=1.75cm,minimum height=0.40cm] at (7.065,1.250) {bootstrap $s_3$};

\node[upl,minimum width=1.00cm,minimum height=0.20cm] at (2.000,2.030) {};
\node[upl,minimum width=1.00cm,minimum height=0.20cm] at (2.000,0.830) {};
\node[upl,minimum width=0.42cm,minimum height=0.20cm] at (3.090,2.030) {};

\draw[->,>=latex,thin] (1.25,0.52) -- (8.30,0.52);
\node[note,anchor=west] at (1.250,0.370) {time (not to scale)};
\node[comp,minimum width=0.35cm,minimum height=0.22cm] at (3.975,0.370) {};
\node[note,anchor=west] at (4.180,0.370) {task};
\node[xchg,minimum width=0.35cm,minimum height=0.22cm] at (5.175,0.370) {};
\node[note,anchor=west] at (5.380,0.370) {exchange};
\node[upl,minimum width=0.35cm,minimum height=0.16cm] at (7.075,0.370) {};
\node[note,anchor=west] at (7.280,0.370) {upload};
\end{tikzpicture}
\caption{A simplified 2-GPU plan for the running example.
(a) Partition and placement after restructuring the projection sums;
edges between rows cross GPUs.
(b) Execution: host-link lanes show streamed plaintext uploads
overlapping computation. Bar lengths are not measurements.}
\label{fig:example-plan}
\end{figure}
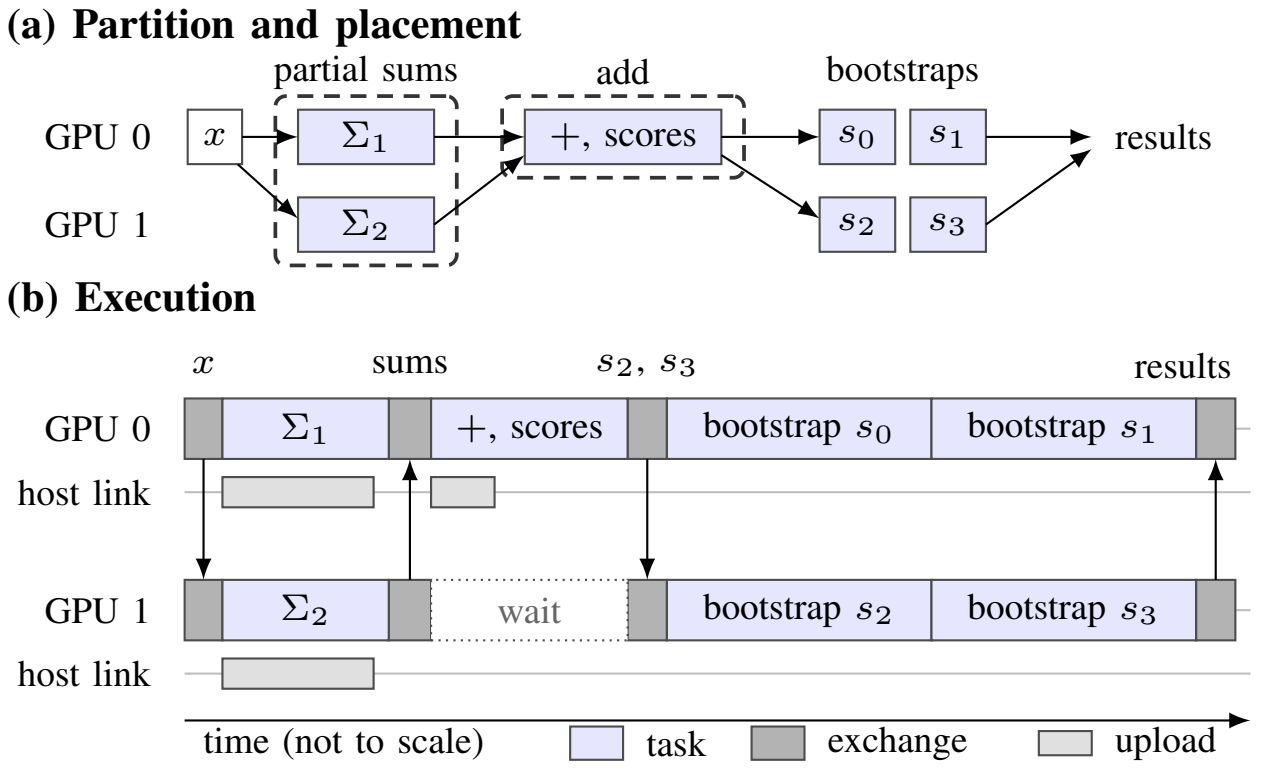

\noindent\textbf{Example.}
In Figure~\ref{fig:example-plan}, the partial-sum region is
split across 2 GPUs; GPU~1 then waits while GPU~0 combines
the sums and computes the scores.
Each GPU streams its assigned plaintext diagonals while
computing partial sums.
Further splitting on GPU~0 could exploit GPU~1's idle time
but would require bidirectional transfers of partial sums.
The planner compares these alternatives using $\hat{T}$.

\subsection{Lowering and Runtime}
\label{sec:runtime}

For the selected plan, the compiler emits a program
for each GPU, specifying local task execution,
inter-GPU transfers, and operand uploads.
Before inference, the runtime captures eligible sequences
of GPU operations into CUDA graphs.
For each request, it replays these graphs and performs
transfers according to the fixed schedule.

\section{Evaluation}
\label{sec:eval}

We evaluate \fwName's performance across workloads and hardware
configurations, compare it with prior multi-GPU FHE libraries,
and measure its compilation overhead.

\subsection{Methodology}
\label{sec:methodology}

\noindent\textbf{Workloads.}
We evaluate \fwName on three representative FHE workloads.
For bootstrapping, we use Cheddar's open-source
implementation~\cite{kim2024cheddar} as the standard benchmark,
with $N=65536$,$\Delta=2^{40}$, 47 primes in $Q$, 12 in $P$,
$\mathrm{dnum}=4$, and $\log PQ\le1747$.
For BERT-Base, we use THOR's open-source
implementation~\cite{moon2025thor}.
For Llama-3-8B prefill, we reimplement
Sylph~\cite{park2026scaling} based on its paper,
as its source code is not publicly available.
Both model workloads retain their original packing schemes,
polynomial approximations, and FHE parameter settings.
The Llama-3-8B workload uses 128 private tokens and 2 public
prefix tokens.
Sylph-based Llama prefill uses early sends. Our THOR-based BERT and
Cerium Llama ablation runs exchange data at the end of each round.

\noindent\textbf{Configurations.}
We evaluate TurboFHE on five 8-GPU systems(Table~\ref{tab:machines}):
RTX~4090 and RTX~5090 with PCIe~4.0~$\times$8 and
PCIe~5.0~$\times$16, respectively;
and A100 SXM4 and H100 SXM with NVSwitch interconnects.

\begin{table}[t]
\centering
\small
\setlength{\tabcolsep}{1.5pt}
\renewcommand{\arraystretch}{1.0}
\caption{Per-GPU hardware specifications.}
\label{tab:machines}
\begin{tabular*}{\columnwidth}{@{\extracolsep{\fill}}lrrrrlr@{}}
\toprule
GPU
& SMs
& \makecell{Mem.\\(GB)}
& \makecell{Mem.\\BW\\(TB/s)}
& \makecell{INT32\\(TOPS)}
& Link
& \makecell{Link\\BW\\(GB/s)} \\
\midrule
RTX 4090
& 128 & 24 & 1.01 & 41.3
& PCIe 4.0 $\times$8 & 15.8 \\
RTX 5090
& 170 & 32 & 1.79 & 104.8
& PCIe 5.0 $\times$16 & 63.0 \\
RTX PRO 6000
& 188 & 96 & 1.60 & 120.0
& PCIe 5.0 $\times$16 & 63.0 \\
A100 SXM
& 108 & 80 & 2.04 & 19.5
& NVSwitch & 300 \\
H100 SXM
& 132 & 80 & 3.35 & 33.5
& NVSwitch & 450 \\
\bottomrule
\end{tabular*}
\end{table}

\subsection{Parallel Efficiency}
Table~\ref{tab:overall-performance} reports execution times across consumer and datacenter GPU systems.
The results show that the best-performing system varies by workload.
The results also show marked disparity in parallel efficiency, ranging from 25.0\% to 69.7\% on 4 GPUs.
These variations highlight the importance of automated, workload- and hardware-aware
planning for multi-GPU FHE execution.
\begin{table}[t]
\centering
\caption{Latency of \fwName on multi-GPUs.}
\label{tab:overall-performance}
\small
\begingroup
\setlength{\tabcolsep}{2pt}
\renewcommand{\arraystretch}{1.12}
\begin{tabular*}{\columnwidth}{@{\extracolsep{\fill}}llrrrrl@{}}
\toprule
\multirow{2}{*}{Workload} & \multirow{2}{*}{GPU} & \multicolumn{4}{c}{\fwName{} latency} & \shortstack{Related work} \\
\cmidrule(lr){3-6}
 & & $1\times$ & $2\times$ & $4\times$ & $8\times$ & 1 GPU \\
\midrule
\multirow{4}{*}{\shortstack[l]{Bootstrap\\(ms)}}
 & 4090 & 33.5 & 33.5 & 33.5 & 33.5 & C: 31.6 \\
 & 5090 & 19.0 & 15.1 & 15.1 & 15.1 & C: 22.1 \\
 & A100 & 31.8 & 23.8 & 19.9 & 18.2 & C: 40.0$^{p}$ \\
 & H100 & 18.8 & 14.7 & 12.6 & 11.5 & C: 31.2$^{p}$ \\
\midrule
\multirow{4}{*}{\shortstack[l]{BERT-Base\\(s)}}
 & 4090 & 26.9 & 20.4 & 15.0 & 12.8 & -- \\
 & 5090 & 17.8 & 11.2 & 8.7 & 8.1 & -- \\
 & A100 & 24.6 & 15.3 & 10.1 & 8.8 & \shortstack[l]{T: 602, A: 745$^{a}$} \\
 & H100 & 15.0 & 9.4 & 6.3 & 4.94 & -- \\
\midrule
\multirow{4}{*}{\shortstack[l]{Llama-3-8B\\prefill (s)}}
 & 4090 & 133.1 & 84.6 & 50.1 & 35.5 & -- \\
 & PRO~6000 & 85.3 & 52.5 & 30.6 & 21.6 & S: 116 \\
 & A100 & 156.7 & 99.7 & 59.2 & 41.2 & -- \\
 & H100 & 94.5 & 59.9 & 35.9 & 25.3 & -- \\
\bottomrule
\end{tabular*}
\par\smallskip
\parbox{\columnwidth}
{\footnotesize
C: Cheddar~\cite{kim2024cheddar};
T: THOR~\cite{moon2025thor};
A: AEGIS~\cite{aegis};
S: Sylph~\cite{park2026scaling}.
$^{p}$PCIe GPU.
$^{a}$A100 40\,GB.
}
\endgroup

\end{table}

\noindent\textbf{Bootstrapping.}
High GPU compute throughput does not guarantee efficient
scaling when communication is costly.
On PCIe-based RTX~4090 and RTX~5090, the planner uses only
one and 2 GPUs, respectively, as further splitting
incurs more communication overhead than computational savings.
Even with NVSwitch, A100 and H100 achieve only
1.75$\times$ and 1.63$\times$ speedups on 8 GPUs,
corresponding to parallel efficiencies of 21.8\% and 20.4\%.
Exposed communication accounts for roughly a quarter
of latency on both systems.
These results favor assigning independent bootstraps to
separate GPUs when sufficient ciphertext-level parallelism
is available.
Accordingly, we initialize the graph with
bootstrapping nodes and maximal connected components
of the remaining ciphertext operations.

\noindent\textbf{BERT-Base.}
On one A100, \fwName achieves a 24.5$\times$ speedup over
THOR's reported latency.
On 8 GPUs, parallel efficiency ranges from 26.3\%
on RTX~4090 to 38.0\% on H100.
PCIe limits bootstrap splitting on RTX GPUs, whereas
NVSwitch enables it on A100 and H100, although critical-path dependencies
and unoverlapped communication still limit scaling.
Kernel profiling attributes most of H100's advantage
over RTX~5090 to inner products accessing large evaluation keys
and plaintext constants. These kernels run 1.43$\times$ faster
on H100, while their DRAM access patterns constrain RTX~5090.

\noindent\textbf{Llama-3-8B prefill.}
Our implementation achieves a relative $L_2$ error of
$3.5\times10^{-4}$ in the decrypted final hidden states.
On one RTX~Pro~6000, \fwName takes 85.3\,s, a 1.36$\times$
speedup over Sylph's reported 116\,s.
On eight RTX PRO~6000 GPUs, \fwName achieves an inference latency of 21.6\,s with an automatically generated plan, approaching the 20\,s reported by Sylph using a specialized multi-GPU implementation.
Across the five platforms, 8-GPU speedups range from
3.74$\times$ to 3.95$\times$, corresponding to parallel
efficiencies of 46.7--49.4\%.
The critical path is dominated by large matrix products in
attention and serial bootstrapping chains in normalization
and softmax.
Finer partitioning can reduce computation time along this path,
but the additional work and transfers largely offset the savings.

\subsection{Ablation Study}
\label{sec:eval-prior}
Using Cerium's open-source Llama-3-8B algorithm~\cite{jayashankar2026cerium} with identical inputs, we compare the multi-GPU scalability of \fwName and Cerium in Table~\ref{tab:llama-scaling}. \fwName achieves higher speedups, reaching $5.56$--$6.34\times$ on 8 GPUs. It outperforms Cerium even on its handwritten plan on H100 platform, with larger latency improvements of $1.14\times$ on A100. These results demonstrate the effectiveness of automated, hardware-aware planning across GPU platforms without manual tuning.
\begin{table}
\centering
\caption{Speedup on one Llama-3-8B block, $T_1/T(P)$.}
\label{tab:llama-scaling}
\footnotesize
\setlength{\tabcolsep}{2pt}
\begin{tabular*}{\columnwidth}{@{\extracolsep{\fill}}llrrrrr@{}}
\toprule
\multirow{2}{*}{Platform} & \multirow{2}{*}{Framework} & \multirow{2}{*}{$T(1)$ (s)}
 & \multicolumn{4}{c}{Speedup at $P$ GPUs} \\
\cmidrule(l){4-7}
 & & & 1 & 2 & 4 & 8 \\
\midrule
\multirow{2}{*}{H100 SXM}
 & \fwName & $10.6135$ & $1.00\times$ & $\mathbf{1.94\times}$ & $\mathbf{3.64\times}$ & $\mathbf{6.23\times}$ \\
 & Cerium & $10.5704$ & $1.00\times$ & $1.91\times$ & $3.53\times$ & $5.64\times$ \\
\addlinespace[2pt]
\multirow{2}{*}{A100 SXM}
 & \fwName & $17.8385$ & $1.00\times$ & $\mathbf{1.96\times}$ & $\mathbf{3.67\times}$ & $\mathbf{6.34\times}$ \\
 & Cerium & $18.2485$ & $1.00\times$ & $1.92\times$ & $3.56\times$ & $5.70\times$ \\
\addlinespace[2pt]
\multirow{2}{*}{RTX PRO 6000}
 & \fwName & $9.5867$ & $1.00\times$ & $\mathbf{1.98\times}$ & $\mathbf{3.66\times}$ & $\mathbf{5.56\times}$ \\
 & Cerium & $10.4894$ & $1.00\times$ & $1.90\times$ & $3.44\times$ & $4.91\times$ \\
\bottomrule
\end{tabular*}
\end{table}

\begin{figure}
\centering
\includegraphics[width=\columnwidth]{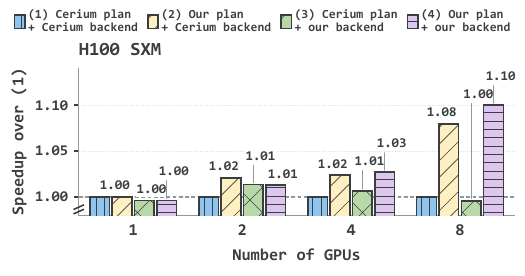}
\caption{Effect of planning and backend on Cerium's Llama-3-8B
workload on H100 SXM GPUs.
Speedup is $T_{(1)}(P)/T(P)$ at GPU count $P$,
where $T$ is per-block latency, and numbers displayed are rounded.
}
\label{fig:llama-ablation}
\end{figure}

We further compare four configurations to isolate planning and execution backend effects
(Figure~\ref{fig:llama-ablation}):
\textbf{C1}: Cerium's handwritten parallelization and backend;
\textbf{C2}: our planner with Cerium's compiler and runtime;
\textbf{C3}: Cerium's handwritten plan with our backend; and
\textbf{C4}: our planner and backend.
In \textbf{C1}, each function declares its GPU count and binds its
streams to fixed GPU groups. Cerium's compiler distributes residue
polynomials across the GPUs in each group~\cite{jayashankar2026cerium}.
We compare per-block latency at $P\in\{1,2,4,8\}$ on H100 GPUs.
At $P=1$, \textbf{C2} is identical to \textbf{C1} and uses its measured latency.

The ablation highlights the importance of backend-aware planning. On 8 H100 GPUs, replacing Cerium's backend while retaining its plan (C3) offers no improvement, with latency 0.4\% higher than C1's latency. In contrast, our planner achieves $1.08\times$ speedup even with Cerium's backend (C2), indicating that planning accounts for most of the gains. Jointly optimizing the plan and backend (C4) further reduces latency by 9.5\% relative to the C3, achieving a $1.10\times$ speedup over C1. The advantage of C4 grows from $1.01\times$ on 2 GPUs to $1.10\times$ on 8 GPUs, demonstrating the increasing value of backend-aware planning as GPU count increases.

\begin{figure}
\centering
\includegraphics[width=\columnwidth]{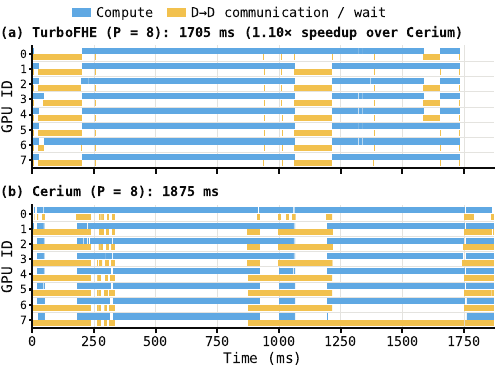}
\caption{Execution timelines for one Llama-3-8B block on 8
H100 GPUs.
}
\label{fig:swimlane-p8}
\end{figure}

Figure~\ref{fig:swimlane-p8} shows kernel-level traces collected
on the same 8-H100 system.
All constants remain resident on their assigned GPUs, so the swimlanes show only computation and
device-to-device communication or waiting.
The timelines include tracing overhead, whereas the latency
and speedup annotations use untraced measurements.
Aggregate computation time is nearly identical:
11.03 GPU-seconds for \fwName and 11.05 for Cerium.
The latency gap therefore reflects differences in load balance and communication.

\textbf{First}, Cerium uses only seven GPUs during the FFN V/W
projections and SiLU stages, leaving GPU~7 idle during these phases
and waiting for 54\% of the full execution.
Our planner balances work using measured task costs,
with per-GPU waiting accounting for 11--23\% of total execution time.

\textbf{Second}, Cerium redistributes function
outputs using all-gather across all P GPUs in the evaluated Llama plan, including GPUs that do not consume those outputs.
Although our planner determines when and between which GPUs
to transfer data, Cerium's collective-based backend does not directly realize the task-specific point-to-point transfers selected by our planner.
In contrast, \fwName uses point-to-point communication between tasks, synchronizng only the sender and receiver.
These collective communications limit further gains
from planning alone.
Together, these results highlight the benefits of jointly optimizing program partitioning, task placement, operand supply, and scheduling.

\subsection{Compilation Cost}
\label{sec:eval-cost}

With generated kernels cached, compiling the DSL computations before
multi-GPU plan search takes 1.0\,s for bootstrapping, 63\,s for one
BERT-Base layer, and 18\,s for one Llama-3-8B layer on an AMD Ryzen 9
9950X. These timings exclude weight encoding.
CUDA graph capture takes 0.17, 2.8, and 2.0\,s, respectively.
Structurally identical layers reuse the compiled program.
For Llama-3-8B, planner search takes 13, 35, and 173\,s
for $P=2,4,8$, respectively.

\section{Conclusion}
We present \fwName, an automated framework for efficient multi-GPU FHE inference. Its computation graph supports splitting at multiple granularities, while a measurement-guided planner jointly optimizes task partitioning, placement, scheduling, and operand residency under hardware constraints. Critical-path rewrites and optimized GPU code generation and runtime further improve execution efficiency. Across diverse workloads and GPU platforms, \fwName automatically generates tailored parallelization plans without manual tuning. On eight GPUs, it achieves $5.56$--$6.34\times$ speedups for a Llama-3-8B block, consistently outperforming Cerium in scalability. These results demonstrate the effectiveness of automated, hardware-aware optimization and the potential of multi-GPU systems for large-scale privacy-preserving inference.

\bibliographystyle{IEEEtran}
\bibliography{reference}

\end{document}